\documentclass[twocolumn,aps,superscriptaddress,showpacs,floatfix,prc,noshowpacs]{revtex4}
\usepackage{mathrsfs}
\usepackage{amssymb}
\usepackage{amsmath}
\usepackage{graphicx}
\usepackage[normalem]{ulem}
\usepackage[dvips]{color}
\usepackage{bm}
\usepackage{longtable}\usepackage{slashed}

\usepackage{enumitem}
\usepackage{empheq}

\usepackage{epstopdf}

\usepackage{fouriernc}

\renewcommand\sout{\bgroup \color{red} \ULdepth=-.5ex \ULset}

\renewcommand{\rm}[1]{\textrm{#1}}
\renewcommand{\d}{\mathrm{d}}

\begin{document}

\title{High-order isospin-dependent surface tension contribution to the fourth-order symmetry energy of finite nuclei}

\author{Bao-Jun Cai\footnote{bjcai87@gmail.com}}
\affiliation{Quantum Machine Learning Laboratory, Shadow Creator Inc., Shanghai 201208, China}
\author{Rui Wang\footnote{wangrui@sinap.ac.cn}}
\affiliation{Key Laboratory of Nuclear Physics and Ion-beam Application (MOE), Institute of
Modern Physics, Fudan University, Shanghai 200433, China}
\author{Zhen Zhang\footnote{zhangzh275@mail.sysu.edu.cn}}
\affiliation{Sino-French Institute of Nuclear Engineering and Technology, Sun Yat-Sen
University, Zhuhai 519082, China}
\author{Lie-Wen Chen\footnote{Corresponding author; lwchen@sjtu.edu.cn}}
\affiliation{School of Physics and Astronomy and Shanghai Key Laboratory for Particle
Physics and Cosmology, Shanghai Jiao Tong University, Shanghai 200240, China}

\date{\today}

\begin{abstract}

The relation between the fourth-order symmetry energy $E_{\rm{sym,4}}(\rho_0)$ of nuclear matter at saturation density $\rho_0$ and its counterpart $a_{\rm{sym,4}}(A)$ of finite nuclei in a semiempirical nuclear mass formula is revisited by considering the high-order isospin-dependent surface tension contribution to the latter.
We derive the full expression of $a_{\rm{sym,4}}(A)$, which includes explicitly the high-order isospin-dependent surface tension effects,
and find that the value of $E_{\rm{sym,4}}(\rho_0)$ cannot be extracted from the measured $a_{\rm{sym,4}}(A)$ before the high-order surface tension is well constrained.
Our results imply that a large $a_{\rm{sym,4}}(A)$ value of several MeVs obtained from analyzing nuclear masses can nicely agree with the empirical constraint of $E_{\rm{sym,4}}(\rho_0)\lesssim 2$~MeV from mean-field models and does not necessarily lead to a large $E_{\rm{sym,4}}(\rho_0)$ value of $\approx 20$~MeV obtained previously without considering the high-order surface tension.
Furthermore,
we also give the expression for the sixth-order symmetry energy $a_{\rm{sym,6}}(A)$ of finite nuclei, which involves more nuclear matter bulk parameters and the higher-order isospin-dependent surface tension.

\end{abstract}

\maketitle

\section{Introduction}\label{S1}

The fourth-order symmetry energy of nuclear matter plays an important role in determining the properties of neutron stars such as the proton fraction and the core-crust transition density/pressure\,\cite{Zha01,Ste06,Xu09,Cai12,Sei14,Gon17,PuJ17}, with the former being critical to neutron star cooling\,\cite{Lat91,Yak01,Yak04}.
Mathematically, the fourth-order symmetry energy of nuclear matter is defined as $E_{\rm{sym,4}}(\rho)\equiv24^{-1} \partial^4E(\rho,\delta)/\partial\delta^4|_{\delta=0}$ where $E(\rho,\delta)$ is the equation of state (EOS) of an asymmetric nucleonic matter (ANM) at density $\rho = \rho_{\rm{n}}+\rho_{\rm{p}}$ and isospin asymmetry $\delta=(\rho_{\rm{n}}-\rho_{\rm{p}})/(\rho_{\rm{n}}+\rho_{\rm{p}})$ with $\rho_{\rm{n}}$($\rho_{\rm{p}}$) denoting the neutron(proton) density\,\cite{Dan02,ditoro,LCK08,Tesym,Col14,Bal16,Oer17,Garg18,LiBA18}.

Most theoretical model calculations indicate that the $E_{\rm{sym,4}}(\rho)$ is much smaller than its preceding term in the ANM EOS, namely, the nuclear symmetry energy defined similarly as $E_{\rm{sym}}(\rho)\equiv2^{-1} \partial^2E(\rho,\delta)/\partial\delta^2|_{\delta=0}$, and this empirical fact can be illustrated from the free Fermi gas (FFG) model.
In particular, in the relativistic FFG model, one has the ratio $\Psi\equiv E_{\rm{sym,4}}(\rho)/E_{\rm{sym}}(\rho)=108^{-1}\cdot(10\nu^4+11\nu^2+4)/(\nu^4+2\nu^2+1)$ with $\nu=k_{\rm{F}}/M$, where $M$ is the nucleon rest mass and $k_{\rm{F}}$ is nucleon Fermi momentum in the symmetric nucleonic matter (SNM).
Consequently, the $\Psi$ takes the value in the range of $1/27\leq\Psi\leq 5/54$\,\cite{CaiLi2022,CaiLi2022a}.
For example,
the $E_{\rm{sym,4}}(\rho_0)$ in the non-relativistic FFG model at the saturation density $\rho_0\approx0.16\,\rm{fm}^{-3}$ is about 0.45\,MeV by taking $M\approx939\,\rm{MeV}$, which is much less than the value of $\sim 13$ MeV for the symmetry energy $E_{\rm{sym}}(\rho_0)$ in the non-relativistic FFG model.
When considering nucleon-nucleon interactions, essentially all the predictions on the fourth-order symmetry energy using either phenomenological models\,\cite{Cai12,Sei14,Gon17,PuJ17,Che09,Agr17} or microscopic many-body theories\,\cite{Lee98,Bom91,Kai15} point to the self-consistent constraint $E_{\rm{sym,4}}(\rho_0)\lesssim2\,\rm{MeV}$.

On the other hand, a systematic expansion (i.e., the ``leptodermous'' expansion) of the energy per nucleon in a finite nucleus is usually made in terms of the small quantity $I\equiv(N-Z)/A\approx\delta$ and the length parameter $A^{-1/3}$, such that $B(N,Z)/A=\sum_{i,j=0,1,2,\cdots}\overline{B}_{ij}I^{2i}A^{-j/3}$ plus terms like the Coulomb and pairing contributions\,\cite{Mye69,Brack85}, here $N$ and $Z$ are the neutron and proton numbers in a finite nucleus with mass number $A=N+Z$, respectively. Specifically, the classic Bethe--Weizs$\ddot{\rm a}$cker mass formula gives\,\cite{Mye69}
\begin{align}
 B(N,Z) =& -a_{\rm{v}}A + a_{\rm{sur}}A^{2/3} + a_{\rm{cou}}\frac{Z^2(1 - Z^{-2/3})}{A^{1/3}}\notag\\
 &+ a_{\rm{a}}I^2A + a_{\rm{p}}\frac{(-1)^{N} + (-1)^Z}{A^{2/3}},
\label{BW}
\end{align}
here $-a_{\rm{v}}$ is the volume energy coefficient, $a_{\rm{sur}}$ is the surface energy coefficient, $a_{\rm{cou}}$ characterizes the Coulomb interaction between protons, $a_{\rm{a}}$ is the so-called symmetry energy coefficient of finite nuclei and $a_{\rm{p}}$ is the paring energy coefficient. By analyzing nuclear masses\,\cite{Mye69}, a few coefficients appearing in Eq.~(\ref{BW}) are found to be about $a_{\rm{v}}\approx15.7\,\rm{MeV},a_{\rm{sur}}\approx18.6\,\rm{MeV}, \rm{and}~ a_{\rm{cou}}\approx0.7\,\rm{MeV}$\,\cite{Mye69}; see also Ref.\,\cite{Mye96}.
By considering the isospin correction to the coefficient of nuclear surface tension [see formula (\ref{st_2})], the symmetry energy $a_{\rm{a}}$ appearing in Eq.~(\ref{BW}) could be generalized to be\,\cite{Dan03}
\begin{equation}\label{finEsym-xx}
a_{\rm{a}}\to a_{\rm{sym}}(A)=\frac{\alpha}{1+(\alpha/\beta)A^{-1/3}},
\end{equation}
which depends only on the mass number $A$.
Here $\alpha=E_{\rm{sym}}(\rho_0)$ and $\beta$ is the so-called surface symmetry energy; for details on $\beta$ see the relevant discussion in Sec.~\ref{S2}.

Generally, even higher order contributions could be considered in the semi-empirical Bethe--Weizs$\ddot{\rm a}$cker mass formula in the form of $a_{\rm{a}}I^2A\to (a_{\rm{a}}I^2+a_{\rm{a,4}}I^4+\cdots)A$, where $a_{\rm{a,4}}$ is the fourth-order symmetry energy coefficient of finite nuclei, which is the main topic of the present work. The $a_{\rm{a,4}}$ coefficient was recently found to have a sizable value by analyzing the double difference of the ``experimental'' symmetry energy extracted from nuclear masses\,\cite{AME2012}.
Specifically, $a_{\rm{a,4}}\approx 3.28\pm0.50$\,MeV was extracted in Ref.\,\cite{Jia14} and $a_{\rm{a,4}}\approx8.33\pm1.21$\,MeV in Ref.\,\cite{Tia16}.
Moreover, in Ref.\,\cite{Jia15}, the fourth-order symmetry energy of finite nuclei was investigated by fitting nuclear mass data via the nuclear mass formula with two different forms of the Winger energy, and the obtained constraint on the $a_{\rm{a,4}}$ is about $3.91\pm0.10\,\rm{MeV}$.

Some questions naturally emerge: What implication does a
sizable $a_{\rm{a,4}}$ give on the value of $E_{\rm{sym,4}}(\rho_0)$?
Is the fourth-order symmetry energy of nuclear matter also large and inconsistent with the empirical constraint that $E_{\rm{sym,4}}(\rho_0)\lesssim2\,\rm{MeV}$?
In this work, the full formula for the fourth-order symmetry energy of finite nuclei is derived, and the conclusion is that although the $a_{\rm{a,4}}$ obtained by fitting nuclear masses is sizable, the corresponding $E_{\rm{sym,4}}(\rho_0)$ could still be relatively small to be consistent with the empirical constraints due to the high-order isospin-dependent surface tension effects.

The paper is organized as follows. Section \ref{S2} gives a brief review of the current status of the symmetry energy and the fourth-order symmetry energy of finite nuclei, starting from the general discussion of Eq.\,(\ref{finEsym-xx}). In Sec.~\ref{S3}, the formula for the fourth-order symmetry energy of finite nuclei is given together with numerical demonstrations, and the emphasis is on the surface contribution.
Section \ref{SS6} is devoted to the analytical expression for the sixth-order symmetry energy of finite nuclei.
Section \ref{S4} gives the summary of the present work.

\section{Symmetry Energies of Finite Nuclei: Relevant Status Review}\label{S2}

We start our discussions by reviewing the status of the symmetry energy of finite nuclei.
For a finite nucleus with $N$ neutrons and $Z$ protons, the total difference $\Delta=N-Z$ could be decomposed into two parts as $\Delta_{\rm{v}}=N_{\rm{v}}-Z_{\rm{v}}$ and $\Delta_{\rm{s}}=N_{\rm{s}}-Z_{\rm{s}}$, representing the isospin difference in the bulk
of the nucleus and that distributed on its surface.
Naturally one has $\Delta=\Delta_{\rm{v}}+\Delta_{\rm{s}}$. As more isospin asymmetry moves to the surface, physically the nucleus becomes looser, indicating that the surface tension of the nucleus becomes smaller compared with the one in which no isospin asymmetry is distributed on the surface. Moreover, as either neutrons or protons are distributed more on the nucleus surface, the surface tension $\sigma=a_{\rm{sur}}/(4\pi r_0^2)$, where $r_0$ is the nuclear matter radius parameter defined by $4\pi\rho_0r_0^3/3 = 1$, always decreases. Consequently, the surface tension with some isospin asymmetry distributed on the nucleus surface could be written as\,\cite{Dan03}
\begin{equation}\label{st_2}
\sigma=\sigma_0-\gamma\mu_{\rm{a}}^2=\sigma_0\left[1-(\gamma/\sigma_0)\mu_{\rm{a}}^2\right],
\end{equation}
where $\mu_{\rm{a}}=(\mu_{\rm{n}}-\mu_{\rm{p}})/2$ is the chemical potential difference between neutrons and protons, and $\gamma$ is a parameter.
The leading non-trivial contribution starting at order $\mu_{\rm{a}}^2$ reflects the aforementioned symmetry between neutrons and protons.

Based on the relation (\ref{st_2}), Ref.\,\cite{Dan03} derived a closed expression for the symmetry energy of finite nuclei incorporating the surface effects, see formula (\ref{finEsym-xx}) where $\alpha\equiv a_{\rm{a}}^{\rm{v}}$ is the coefficient in front of $(N_{\rm{v}}-Z_{\rm{v}})^2/A$, and $\beta\equiv1/16\pi r_0^2\gamma\equiv a_{\rm{a}}^{\rm{s}}$.
The $a_{\rm{a}}^{\rm{v}}$ and $a_{\rm{a}}^{\rm{s}}$ are the volume (bulk) symmetry energy and the surface symmetry energy\,\cite{Dan03,Dan09}, respectively.
In the limit of large $A$, the isospin asymmetry gets primarily stored within the bulk and the
coefficient $a_{\rm{sym}}(A)$ tends towards $a_{\rm{a}}^{\rm{v}}$, i.e.,
we also have $a_{\rm{a}}^{\rm{v}}=E_{\rm{sym}}(\rho_0)$. On the other hand, in the limit of small mass number $A$, the storage of the isospin
asymmetry gets shifted to the surface and the ratio $a_{\rm{sym}}(A)/A$ scales as $a_{\rm{a}}^{\rm{s}}/A^{2/3}$.

In addition, the relation is further established between the coefficient $\beta$ and the ones appearing in the nuclear droplet model\,\cite{Mye69}, i.e.,
$\beta={4Q}/{9}=({4H}/{9})/(1-{2P}/{3J})$,
where $J\equiv E_{\rm{sym}}(\rho_0)=\alpha$, $Q$ is the neutron skin stiffness coefficient\,\cite{Mye96} and the individual constants $H, P$ and $G$, found equal to $G = 3JP/2Q$, describe the dependence of the surface energy on the bulk isospin asymmetry and on normalized size of the neutron skin.

The uncertainties on the symmetry energy of finite nuclei are mainly due to those on the surface symmetry energy coefficient $\beta$, e.g., Ref.\,\cite{Dan03} constrained the ranges for the parameters $\alpha$ and $\beta$ as 27\,MeV$\lesssim\alpha\lesssim$31\,MeV and $11\,\rm{MeV}\lesssim\beta\lesssim14\,\rm{MeV}$ and their ratio as $2.0\lesssim\alpha/\beta\lesssim2.8$\,\cite{Dan03}.
Similarly, an earlier analysis via the Thomas--Fermi model gives $Q$ about 35.4\,MeV, leading to the $\beta$ about 15.7\,MeV, and even more earlier the droplet model gave $Q\approx16\,\rm{MeV}$ and $\beta\approx 7.1\,\rm{MeV}$\,\cite{Mye69}.
Recently, based on analysis of the isobaric analog states (IAS), the volume symmetry energy coefficient and the surface symmetry energy coefficient were found to be about $a_{\rm{a}}^{\rm{v}}=\alpha\approx35.3\,\rm{MeV}$ and $a_{\rm{a}}^{\rm{s}}=\beta\approx 9.7\,\rm{MeV}$, respectively\,\cite{Dan14}, and consequently $\alpha/\beta\approx3.6$.
On the other hand,
when combining the IAS analysis and the neutron skin thickness ($\Delta r_{\rm{np}}$) constraint, the relation between the coefficients $a_{\rm{a}}^{\rm{v}}$ and $a_{\rm{a}}^{\rm{s}}$ is found to be slightly changed. Specifically, the $a_{\rm{a}}^{\rm{v}}$ and the $a_{\rm{a}}^{\rm{s}}$ obtained in the half-infinite calculation are found to be about 30.2-33.7\,MeV and 14.8-18.5\,MeV, leading to the ratio $\alpha/\beta\approx1.92\pm0.24$, while the best values from the combined analysis from IAS and the $\Delta r_{\rm{np}}$ are about 33.2\,MeV and 10.7\,MeV, respectively\,\cite{Dan14}, and thus $\alpha/\beta\approx3.1$.

We have made attempt to review all the relevant investigations of the coefficients $a_{\rm{a}}^{\rm{v}}$ and $a_{\rm{a}}^{\rm{s}}$. The main point we would like to stress here is that these uncertainties may further induce essential uncertainties on the fourth-order symmetry energy of finite nuclei through the ratio $\alpha/\beta$.

It should be pointed out that if the nuclear surface tension is truncated as in Eq.~(\ref{st_2}), there would be no higher order symmetry energies from the surface contribution in finite nuclei.
However, it is natural and physical that the nuclear surface tension should include higher order contributions, i.e.,
\begin{equation}\label{st-kk}
\sigma\approx\sigma_0-\gamma\mu_{\rm{a}}^2-\gamma'\mu_{\rm{a}}^4-\gamma''\mu_{\rm{a}}^6+\cdots,
\end{equation}
where $\gamma'$ and $\gamma''$, etc., are effective parameters.
Under this situation, one can obtain higher order terms in the binding energy per nucleon in finite nuclei, which is the starting point for the calculations of the present work.

Recently, the bulk term $a_{\rm{sym,4}}^{\rm{v}}(A)$ of the fourth-order symmetry energy $a_{\rm{sym,4}}(A)$ of finite nuclei was derived using a variational method in Ref.\,\cite{WangRui17} based on the lowest-order isospin truncation of Eq.\,(\ref{st_2}) for the surface tension, and the analytic expression is obtained as
\begin{equation}\label{a_a4v}
a_{\rm{sym,4}}^{\rm{v}}(A)=\frac{1}{(1+\alpha/\beta A^{1/3})^4}\left(E_{\rm{sym},4}(\rho_0)-\frac{L^2}{2K_0}\right).
\end{equation}
Based on this formula, Ref.\,\cite{WangRui17} claimed that the $E_{\rm{sym,4}}(\rho_0)$ should be as large as about $20\,\rm{MeV}$ if $a_{\rm{a,4}}\approx3.28\,\rm{MeV}$ is used\,\cite{Jia14}. Such a large value of $E_{\rm{sym,4}}(\rho_0)\approx 20$ MeV is partially due to the term $L^2/2K_0$ which is about 4.2\,MeV if $L\approx45\,\rm{MeV}$\,\cite{LiBA13,Zha13,Oer17} and $K_0\approx240\,\rm{MeV}$\,\cite{You99,Shl06,Che12,Garg18} are adopted,
and this confusing value is significantly larger than the empirical constraint of $E_{\rm{sym,4}}(\rho_0)\lesssim2\,\rm{MeV}$.
In the present work,
we show that a high-order surface contribution will appear in the fourth-order symmetry energy of finite nuclei due to the high-order isospin dependent surface tension ($\gamma'$) as in Eq.\,(\ref{st-kk}),
and thus
a large $a_{\rm{a,4}}$ value of several MeVs obtained from analyzing nuclear masses does not have to lead to a large $E_{\rm{sym,4}}(\rho_0)$ value of $\approx 20$~MeV obtained without considering the high-order isospin dependent surface tension.

\section{Fourth-order Symmetry Energy of Finite Nuclei: Full Expression}\label{S3}

Generally, one can obtain the following equations to determine the bulk and surface isospin asymmetries as well as the chemical potential difference in finite nuclei\,\cite{Dan03},
\begin{equation}\label{ther33}
\Delta_{\rm{v}}+\Delta_{\rm{s}}=\Delta,~~2\alpha \Delta_{\rm{v}}/A-\mu_{\rm{a}}=0,~~\Delta_{\rm{s}}/S+\d\sigma/\d\mu_{\rm{a}}=0,
\end{equation}
where $S=4\pi r_0^2A^{2/3}$ is the surface area of the nucleus via the relation $R=r_0A^{1/3}$.
It should be pointed out that the equations shown in (\ref{ther33}) are not complete in the sense that as higher order terms like the $\gamma'\mu_{\rm{a}}^4$ would induce a high order symmetry energy in finite nuclei related to the surface properties, one needs to add the bulk term $4\alpha_4(\Delta_{\rm{v}}/A)^3\equiv 4E_{\rm{sym,4}}(\rho_0)(\Delta_{\rm{v}}/A)^3$ originating from the fourth-order symmetry energy to the second equation and to solve self-consistently. However, since this part was already obtained in Ref.\,\cite{WangRui17}, for the purpose of the present work, we will not give the detailed derivations here and only focus on the surface contribution.

\subsection{Effective Symmetry Energies for Finite Nuclei}

By introducing the function $f=\sigma/\sigma_0$, one obtains in the situation $f=1-\theta\mu_{\rm{a}}^2\equiv 1+y$ (with $\theta\equiv\gamma/\sigma_0$ and $y\equiv -\theta\mu_{\rm{a}}^2$) the following expressions for $\Delta_{\rm{v}},\Delta_{\rm{s}}$ and $\mu_{\rm{a}}$,
\begin{equation}\label{exex}
\frac{\Delta_{\rm{v}}}{\Delta}=\frac{1}{1+\phi},~~\frac{\Delta_{\rm{s}}}{\Delta}=\frac{\phi}{1+\phi},~~\mu_{\rm{a}}=\frac{2\alpha}{A}\frac{\Delta}{1+\phi},
\end{equation}
where $\phi=\alpha/\beta A^{1/3}$.
The effective symmetry energy (appearing in the mass formula in the form of $a_{\rm{a}}^{\rm{eff}}(N,Z)I^2A$) in finite nuclei could be obtained as,
\begin{equation}
a_{\rm{a}}^{\rm{eff}}(N,Z)=\left.\left[\frac{\alpha \Delta_{\rm{v}}^2}{A}+\mu_{\rm{a}}\Delta_{\rm{s}}+\sigma_0S(f-1)\right]\right/AI^2,
\end{equation}
which includes the effects from higher order symmetry energies as $a_{\rm{a}}^{\rm{eff}}(N,Z)\approx a_{\rm{sym}}(A)+a^{\rm s}_{\rm{sym,4}}(A)I^2+\cdots$.
In particular, for $f=1-\theta\mu_{\rm{a}}^2$ the effective symmetry energy $a_{\rm{a}}^{\rm{eff}}(N,Z)$ reduces to $a_{\rm{sym}}(A)=\alpha/(1+\alpha/\beta A^{1/3})$.
Moreover, in this simple model there is only one effective parameter $\theta$, and it is determined uniquely by the surface symmetry energy coefficient $\beta$ and the surface tension $\sigma_0$.
Similarly,
the effective fourth-order symmetry energy of finite nuclei is defined as
\begin{equation}\label{oka4}
a_{\rm{a,4}}^{\rm{eff}}(N,Z)=[{a_{\rm{a}}^{\rm{eff}}(N,Z)-a_{\rm{sym}}(A)}]/{I^2},
\end{equation}
which contains even higher order contributions, e.g., the sixth-order symmetry energy.

As long as the function $f$ is depending on the combination $y=-\theta\mu_{\rm{a}}^2$, i.e., $f=f(y)$, it could be proved straightforwardly that the bulk and the surface isospin asymmetries are given, by generalizing the first two relations of (\ref{exex}), as
\begin{align}
\frac{\Delta_{\rm{v}}}{\Delta}=&\left(1+\phi
\frac{\d f}{\d y}\right)^{-1},~~
\frac{\Delta_{\rm{s}}}{\Delta}=\phi\frac{\d f}{\d y}\left(1+\phi
\frac{\d f}{\d y}\right)^{-1}.\label{ckck-12}
\end{align}
In this situation, the effective symmetry energy of finite nuclei can be obtained as
\begin{align}
a_{\rm{a}}^{\rm{eff}}(N,Z)
=&\frac{\alpha(1+2\phi
{\d f}/{\d y})}{(1+\phi
{\d f}/{\d y})^2}
+\frac{4\pi r_0^2\sigma_0}{I^2A^{1/3}}\left[f(y)-1\right],\label{aNZ}
\end{align}
and the $\mu_{\rm{a}}$ should be self-consistently obtained by solving the three equations of (\ref{ther33}).
As $A\to\infty$ and $N-Z\to\infty$ but the ratio $I$ is fixed, the chemical potential difference $\mu_{\rm{a}}$ will also be fixed for a given $I$. In particular, all effective models for $f$ tend to be the same in the large-$A$ limit and $\mu_{\rm{a}}\approx\mu_{\rm{a}}^{\infty}\equiv2\alpha I$, or equivalently $2\mu_{\rm{a}}=\mu_{\rm{n}}-\mu_{\rm{p}}\approx 4\alpha\delta$, indicating that the $a_{\rm{a,4}}^{\rm{eff}}(N,Z)$ approaches to zero in this limit. It should be remembered that only the surface contribution to the fourth-order symmetry energy is studied here, see the comments given at the beginning of this section.
If $f=1+y$ is adopted, then Eq.\,(\ref{aNZ}) naturally reduces to $\alpha/(1+\phi)$.

\subsection{Surface Fourth-order Symmetry Energy}

Formula (\ref{aNZ}) itself could be used to derive the fourth-order symmetry energy of finite nuclei.
We start from formula (\ref{aNZ}) by assuming that the function $f$ takes the form $
f\approx1+y+\kappa y^2$ where $\kappa=-\gamma'\sigma_0/\gamma^2$ is an effective parameter characterizing the fourth-order contribution to the isospin splitting of the $f=\sigma/\sigma_0$ [see Eq.~(\ref{st-kk})]. Consequently one has $f'\equiv\d f/\d y=1+2y\kappa$, and the first term in Eq.~(\ref{aNZ}) becomes
\begin{align}\label{dk-1}
\frac{\alpha(1+2\phi f')}{(1+\phi f')^2}\approx\frac{\alpha(1+2\phi)}{(1+\phi)^2}-\frac{4\phi^2y\alpha}{(1+\phi)^3}\kappa,
\end{align}
to order $\kappa$.
Similarly the second term in Eq.~(\ref{aNZ}) is expanded as $
{4\pi r_0^2\sigma_0(y+y^2\kappa)}/{I^2A^{1/3}}$,
where the first term here gives
\begin{equation}\label{dk-2}
\frac{4\pi r_0^2\sigma_0y}{I^2A^{1/3}}\approx-\frac{\alpha\phi}{(1+\phi)^2}-\frac{16\alpha^3\phi^2\theta I^2\kappa}{(1+\phi)^5}.
\end{equation}
Summing the first term of Eq.~(\ref{dk-1}) and the first term of Eq.~(\ref{dk-2}) gives the familiar formula for symmetry energy of finite nuclei, i.e., $a_{\rm{sym}}(A)=\alpha/(1+\phi)$.

Since we assume $\kappa$ is small, the $y$ in $4\pi r_0^2\sigma_0y^2\kappa/I^2A^{1/3}$ could be approximated as
$y\approx-{4\theta\alpha^2I^2}/{(1+\phi)^2}$ (recalling that $\Delta/A=I$),
and moreover $
\mu_{\rm{a}}\approx{2\alpha I}/({1+\phi})$.
One then obtains
\begin{align}
-\frac{4\alpha\phi^2y}{(1+\phi)^3}\approx&\frac{16\alpha^3\phi^2\theta}{(1+\phi)^5}I^2,~~
\frac{4\pi r_0^2\sigma_0y^2}{I^2A^{1/3}}\approx\frac{4\alpha^3\phi\theta }{(1+\phi)^4}I^2,
\end{align}
where the relation
$
{4\pi r_0^2\sigma_0y}/{I^2A^{1/3}}\approx-{\alpha\phi}/{(1+\phi)^2}
$
is used for obtaining ${4\pi r_0^2\sigma_0y^2}/{I^2A^{1/3}}$ at this order.

Combining all these terms gives the effective symmetry energy of finite nuclei to order $I^2\kappa$ as
\begin{equation}
a_{\rm{a}}^{\rm{eff}}(N,Z)\approx \frac{\alpha}{1+\phi}
+\frac{4\alpha^3\phi\theta \kappa I^2}{(1+\phi)^4},
\end{equation}
the coefficient in front of $I^2$ of the second term gives
the (surface) fourth-order symmetry energy of finite nuclei, i.e.,
\begin{equation}\label{a4ext}
a^{\rm s}_{\rm{sym,4}}(A)=\frac{4\alpha^3\theta \kappa\phi}{(1+\phi)^4}
=\left.\frac{4\alpha^4\theta\kappa}{\beta A^{1/3}}\right/\left(1+\frac{\alpha}{\beta A^{1/3}}\right)^4.
\end{equation}
By combining Eq.~(\ref{a4ext}) with the bulk contribution Eq.~(\ref{a_a4v}) derived in Ref.\,\cite{WangRui17}, we finally obtain the total fourth-order symmetry energy of finite nuclei as
\begin{align}\label{def_asym4}
a_{\rm{sym,4}}(A)=&\left(1+\frac{E_{\rm{sym}}(\rho_0)}{\beta A^{1/3}}\right)^{-4}\notag\\
&\times\left(E_{\rm{sym,4}}(\rho_0)
-\frac{L^2}{2K_0}+\frac{4\theta\kappa E_{\rm{sym}}^4(\rho_0)}{\beta A^{1/3}}\right).
\end{align}

\begin{figure}[h!]
\centering
\includegraphics[height=4.5cm]{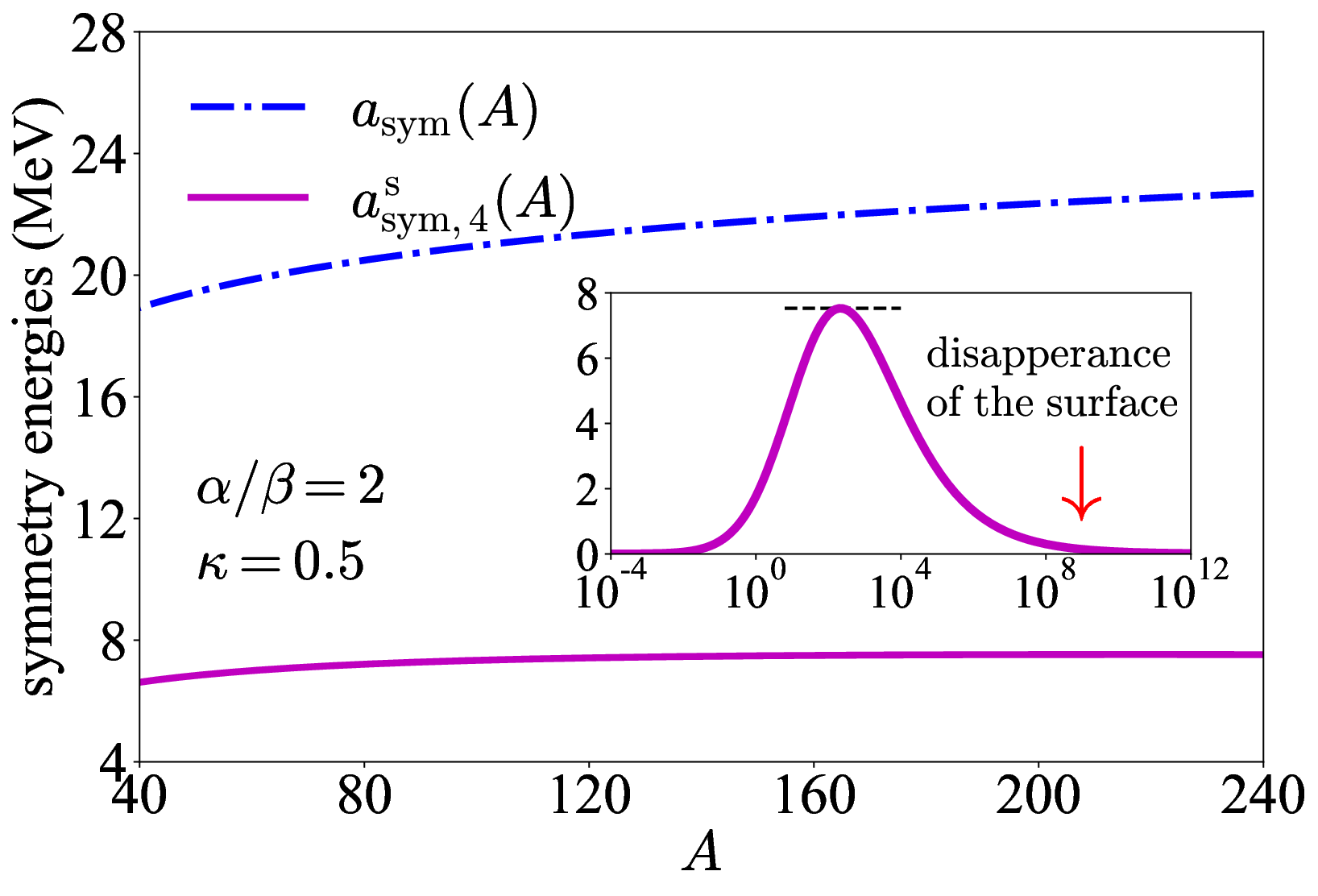}
\caption{The symmetry energy $a_{\rm{sym}}(A)$ and the surface fourth-order symmetry energy $a^{\rm s}_{\rm{sym,4}}(A)$ of finite nuclei as functions of nuclear mass number $A$ . Here $\alpha/\beta=2$ and $\kappa=0.5$ are adopted for illustration.}
\label{fig_a4exact}
\end{figure}

In Fig.\,\ref{fig_a4exact}, the $A$ dependence of the symmetry energy $a_{\rm{sym}}(A)$ (blue dash-dot line) and the surface fourth-order symmetry energy $a^{\rm s}_{\rm{sym,4}}(A)$ (magenta line) of finite nuclei are shown by adopting $\kappa=0.5$, and moreover $\alpha\approx 30\,\rm{MeV},
\beta\approx 15\,\rm{MeV},r_0\approx 1.12\,\rm{fm}$\,\cite{Dan03}, and $
\sigma_0\approx0.8\,\rm{MeV}/\rm{fm}^2$ are used for illustration.
It is found that the fourth-order symmetry energy (due to the surface) has a weak dependence on the mass number $A$ within the given range. In addition, we have
$
a_{\rm{sym}}(A)\to\beta A^{1/3}, a^{\rm s}_{\rm{sym},4}(A)\to4\kappa\theta\beta^3A$, and $a^{\rm s}_{\rm{sym},4}(A)/a_{\rm{sym}}(A)\to 4\kappa\theta\beta^2A^{2/3}$ as $A\to0$,  and all are independent of the bulk term $\alpha$ and approach to zero as $A\to0$.
For heavy nuclei the $\phi=\alpha/\beta A^{1/3}$ is generally smaller than unity since the mass number $A$ is large, leading to the approximation
$
a^{\rm s}_{\rm{sym,4}}(A)\approx4\alpha^3\theta\kappa\phi=
{\alpha^3A^{2/3}\kappa\phi}/{S\beta\sigma_0}$.
The interesting feature of this approximated fourth-order symmetry energy is that it is linearly proportional to the factor $\phi$ which approaches to zero in infinite matter limit, i.e., $
\lim_{A\to\infty}a^{\rm s}_{\rm{sym,4}}(A)=0\,\rm{MeV}$.
There is no surprise that $a^{\rm s}_{\rm{sym},4}(A)$ approaches to zero as $A\to\infty$ since it only reflects the surface part of the fourth-order symmetry energy of finite nuclei.
Physically the surface disappears as the mass number $A$ approaches to infinity, i.e., the related fourth-order symmetry energy becomes zero at this limit, see the inset of Fig.\,\ref{fig_a4exact}.
However, it does not mean that the fourth-order symmetry energy for infinite matter should be zero, as in the above calculations the relevant term related to $E_{\rm{sym,4}}(\rho_0)$ is not included in the second equation of (\ref{ther33}), i.e., the fourth-order symmetry energy of finite nuclei obtained here is still characterized by the symmetry energy coefficients $\alpha$ and $\beta$ instead of the $E_{\rm{sym,4}}(\rho_0)$, see Ref.\,\cite{WangRui17} for the relevant discussions.
Nonetheless, (\ref{a4ext}) is enough for our purpose.

Moreover, the value of $A$ corresponding to the maximum of $a^{\rm s}_{\rm{sym},4}(A)$ could be found via $\partial a^{\rm s}_{\rm{sym},4}(A)/\partial A=0$, and this gives $A_{\max}=27\alpha^3/\beta^3\approx216$. Consequently, $a^{\rm s}_{\rm{sym},4}(A_{\max})=27\alpha^3\theta\kappa/64\approx7.5\,\rm{MeV}$; see the black dashed line of the inset in Fig.\,\ref{fig_a4exact}.
We find that the empirical ratio $\alpha/\beta$ (near 2-3) coincidentally predicts that the fourth-order symmetry energy (due to the surface contribution) maximizes near the $^{208}\rm{Pb}$. Infinite matter (with $A\to\infty$) and finite nuclei (with $A$ being around 208) are fundamentally different from this perspective, and this explains the confusion given in Ref.\,\cite{WangRui17}.

The relation $2\alpha\Delta_{\rm{v}}/A=\mu_{\rm{a}}$ could itself be solved perturbatively order by order.
Under the assumption $f=1+y+\kappa y^2$, one obtains the equation for determining the chemical potential using the expression for $\Delta_{\rm{v}}/\Delta$ [see the relations (\ref{ckck-12})],
\begin{equation}\label{mua-xxx}
\mu_{\rm{a}}+\phi\mu_{\rm{a}}-2\kappa\phi\theta\mu_{\rm{a}}^3=2\alpha I.
\end{equation}
In the infinite matter limit the $\phi\to0$ and the equation gives $\mu_{\rm{a}}=\mu_{\rm{a}}^{\infty}\equiv 2\alpha I$.
By treating both the $\phi$ and $\kappa$ perturbatively on the same order, one writes down the $\mu_{\rm{a}}$ to the second order as $
\mu_{\rm{a}}\approx \mu_{\rm{a}}^{\infty}(1+\varphi_1\phi+\varphi_2\kappa+\varphi_3\phi\kappa+\varphi_4\phi^2+\varphi_5\kappa^2)$,
with $\varphi_1-\varphi_5$ five coefficients to be determined. Since there are already two small quantities $\kappa\phi$ in front of $\mu_{\rm{a}}^3$ one could safely approximate the cube of $\mu_{\rm{a}}^3$ as $\mu_{\rm{a}}^{\infty,3}$ to second order. Equation\,(\ref{mua-xxx}) becomes
\begin{align}
&\mu_{\rm{a}}^{\infty} \left(1+\varphi_1\phi+\varphi_2\kappa+\varphi_3\phi\kappa+\varphi_4\phi^2+\varphi_5\kappa^2\right)\notag\\
&+\mu_{\rm{a}}^{\infty}\phi \left(1+\varphi_1\phi+\varphi_2\kappa\right)-2\kappa\phi\theta\mu_{\rm{a}}^{\infty,3}=2\alpha I,
\end{align}
which could be calculated order by order: 1) at zeroth order: one has the result $\mu_{\rm{a}}^{\infty}\approx2\alpha I$, or $\mu_{\rm{n}}^{\infty}-\mu_{\rm{p}}^{\infty}\approx 4\alpha I $; 2) at the first order: we have  $\mu_{\rm{a}}^{\infty}[(\varphi_1+1)\phi+\varphi_2\kappa]=0$, consequently $\varphi_1=-1,\varphi_2=0$; and 3) at second order: one has the equation $\mu_{\rm{a}}^{\infty}[\varphi_3\phi\kappa+\varphi_4\phi^2+\varphi_5\kappa^2+\varphi_1\phi^2+\varphi_2\phi\kappa-2\kappa\phi\theta\mu_{\rm{a}}^{\infty,2}]=0$, and solving it gives $\varphi_4=1,\varphi_5=0$, and $\varphi_3=2\theta\mu_{\rm{a}}^{\infty,2}$.

By combining the above results one obtains the chemical potential difference between neutrons and the protons in finite nuclei to the second order as
\begin{equation}\label{mua-xyz-1}
\mu_{\rm{a}}\approx\mu_{\rm{a}}^{\infty}\left(1-\phi+\phi^2+2\kappa\phi\theta\mu_{\rm{a}}^{\infty,2}\right).
\end{equation}
The relation (\ref{mua-xyz-1}) could be written in the slightly different form, $
\mu_{\rm{n}}-\mu_{\rm{p}}
\approx4\alpha I(1-\phi+\phi^2)+4\kappa\phi\theta\mu_{\rm{a}}^{\infty,3}=4\alpha I(1-\phi+\phi^2)+32\alpha^3\kappa\phi\theta I^3$, via the identity $2\mu_{\rm{a}}=\mu_{\rm{n}}-\mu_{\rm{p}}$.
Moreover, near the infinite matter limit we have $a^{\rm s}_{\rm{sym},4}(A)\approx 4\alpha^3\theta\kappa\phi$ (the surface contribution), and one finally obtains
$
\mu_{\rm{n}}-\mu_{\rm{p}}
\approx4I\alpha (1-\phi+\phi^2)+8I^3a^{\rm s}_{\rm{sym},4}(A)$.
It is analogous to the relation $\mu_{\rm{n}}-\mu_{\rm{p}}\approx4\delta E_{\rm{sym}}(\rho)+8\delta^3E_{\rm{sym,4}}(\rho)$ often used in determining the proton fraction $x_{\rm{p}}=(1-\delta)/2$ in neutron stars\,\cite{Cai12,Gon17,PuJ17}.
In fact
a more similar relation can be found for the neutron and proton chemical potential difference, i.e., $\mu_{\rm{n}}-\mu_{\rm{p}}\approx 4I a_{\rm{sym}}(A)[1+2I^2a^{\rm s}_{\rm{sym},4}(A)/a_{\rm{sym}}(A)]$ in finite nuclei, by recalling that $1-\phi+\phi^2\approx1/(1+\phi)$ and $\alpha(1-\phi+\phi^2)\approx a_{\rm{sym}}(A)$.

\subsection{Implications of the Smallness of $E_{\textmd{sym},4}(\rho_0)$}

Since the derivation of the term $4\theta\kappa E_{\rm{sym}}^4(\rho_0)/\beta A^{1/3}$ is independent of the bulk contribution $E_{\rm{sym,4}}(\rho_0)-L^2/2K_0$, the two terms are additive. Here the $\theta\kappa$-term in Eq.~(\ref{def_asym4}) approaches to zero as $A\to\infty$ since the surface disappears for infinite matter. On the other hand, the bulk term approaches to a constant (independent of $A$), i.e., $E_{\rm{sym,4}}(\rho_0)-L^2/2K_0$ as $A\to\infty$.
It is now clear that one could still have a $E_{\rm{sym,4}}(\rho_0)\lesssim2\,\rm{MeV}$ to be consistent with microscopic calculations, irrespective of the value of $a^{\rm s}_{\rm{sym},4}(A)$ since the surface contribution only affects the finite nuclei.
Solving Eq.~(\ref{def_asym4}) for $E_{\rm{sym,4}}(\rho_0)$ gives the expression for $E_{\rm{sym},4}(\rho_0)$, from which it is clearly demonstrated that a large value of $a_{\rm{a,4}}$ (as obtained from nucleus mass formula fitting) should not necessarily lead to a large $E_{\rm{sym,4}}(\rho_0)$, and the balance strongly depends on the higher order coefficient $\kappa$, which has very little influence on nuclear structure quantities such as the surface tension of certain typical finite nuclei.

\begin{figure}[htb]
\centering
\includegraphics[height=3.8cm]{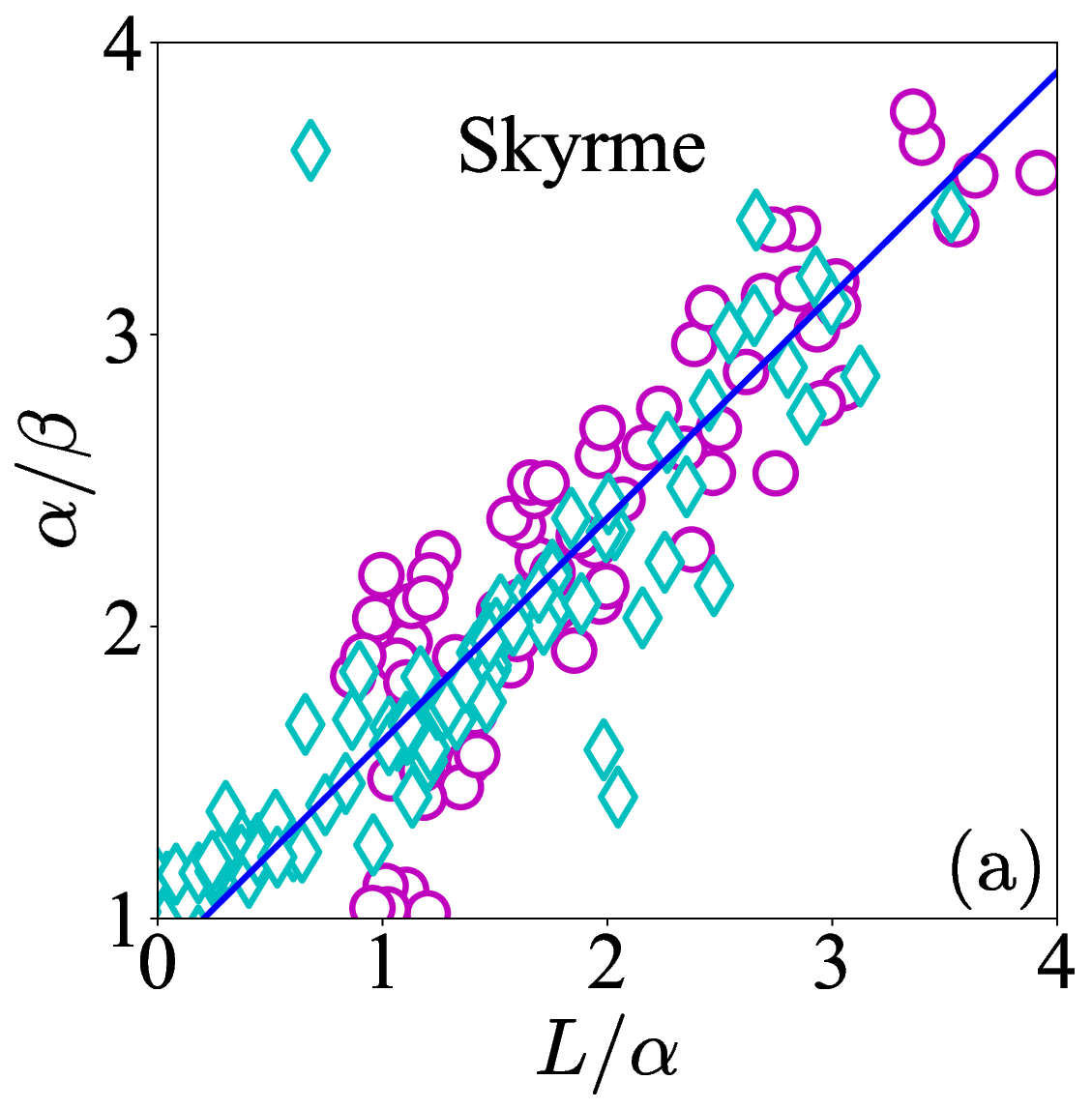}\quad
\includegraphics[height=3.8cm]{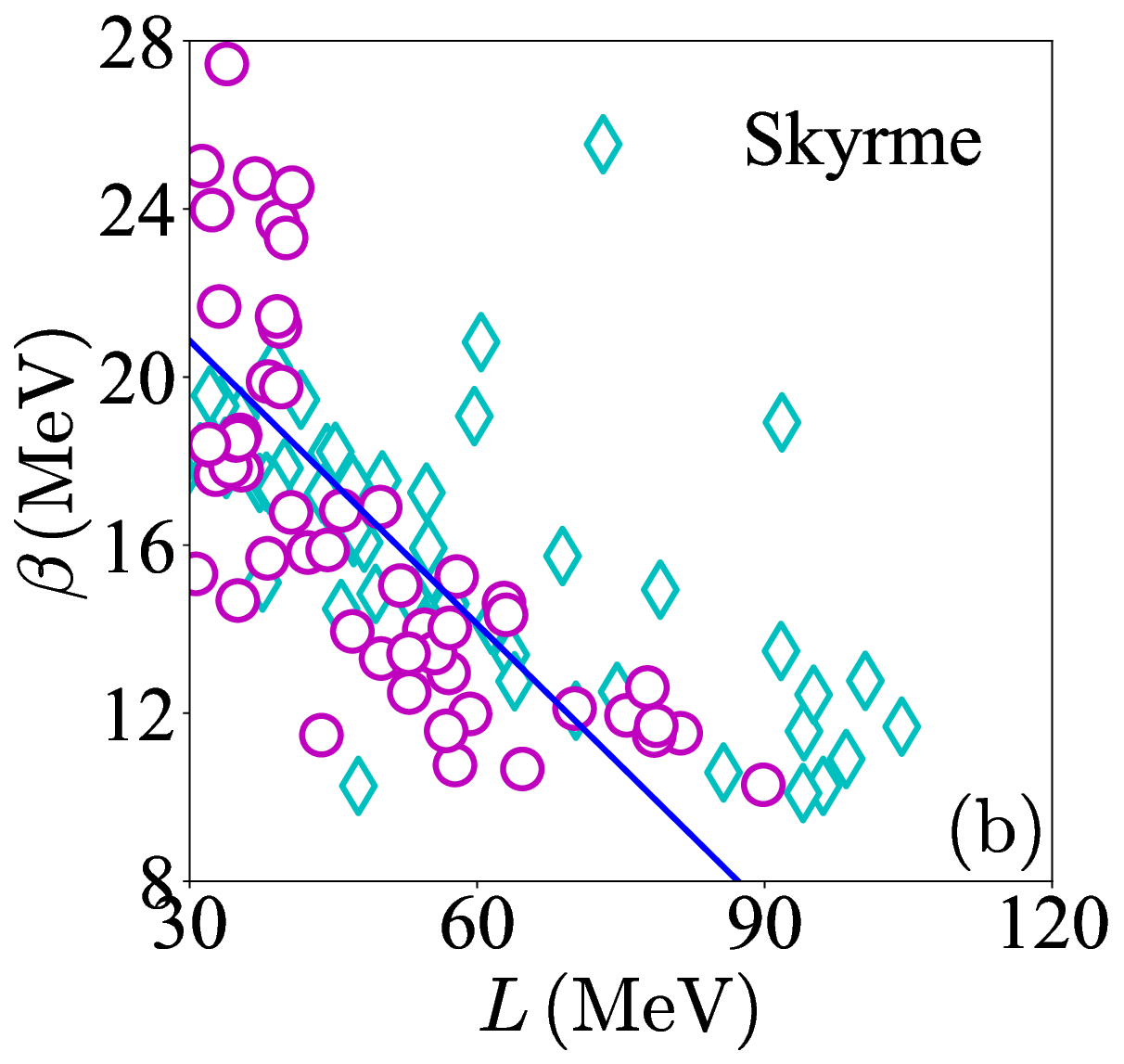}
\caption{The correlation between $\alpha/\beta$ and $L/\alpha$ (left) and between $\beta$ and $L$ (right) via the smallness of $E_{\rm{sym},4}(\rho_0)$. See the text for details.}
\label{fig_ab-La}
\end{figure}

If one accepts the fact that $E_{\rm{sym},4}(\rho_0)$ is empirically smaller than about 2\,MeV\,\cite{Cai12,Gon17,PuJ17}, then Eq.~(\ref{def_asym4}) gives certain correlations among quantities with sizable magnitude.
In this sense, the resulting correlations are expected to be intrinsic like those obtained from the unbound nature of a pure neutron matter\,\cite{Cai21}.
As an example, we study the correlation between the surface symmetry energy coefficient $\beta$ and the slope parameter $L$, by uniformly sampling within the empirical ranges for the symmetry energy as $\alpha=E_{\rm{sym}}(\rho_0)\approx28-36\,\rm{MeV}$\,\cite{Chen17,LiBA21}, the ratio $\alpha/\beta\approx1-4$, the nucleon surface tension $\sigma_0\approx0.6-1.0\,\rm{MeV}/\rm{fm}^2$ in SNM, the coefficient $\kappa\approx0-1.0$, the slope parameter $L\approx30-120\,\rm{MeV}$ of the symmetry energy, the incompressibility coefficient $K_0\approx220-260\,\rm{MeV}$ of SNM, the fourth-order symmetry energy $a_{\rm{a},4}\approx2.78-3.78\,\rm{MeV}$\,\cite{Jia14} extracted from nuclear mass data, $E_{\rm{sym},4}(\rho_0)\approx0-2\,\rm{MeV}$, $a_{\rm{cou}}\approx0.6-0.8\,\rm{MeV}$, and $r_0\approx1.12\,\rm{fm}$.
The model $f$ is simply truncated as $f\approx1+y+\kappa y^2$.
The results calculated
from the samples using Eq.~(\ref{def_asym4}) are then shown as open circles in Fig.\,\ref{fig_ab-La}. For comparison, the results from various Skyrme interactions reported in Ref.\,\cite{Dan09} are present as open diamonds.
Specifically, the correlation of the ratio $\alpha/\beta$ to the ratio $L/\alpha$ is shown in the left panel of Fig.\,\ref{fig_ab-La}, which demonstrates well linear dependence.
This is consistent with the findings using the Skyrme interactions\,\cite{Dan09}.
Similarly, the correlation between $\beta$ and $L$ is shown in the right panel of Fig.\,\ref{fig_ab-La}.

Furthermore, if a macroscopic formula for the neutron skin thickness $\Delta r_{\rm{np}}$ for heavy nuclei is adopted, then one can similarly investigate the correlations between $\Delta r_{\rm{np}}$ and  characteristic parameters such as the surface symmetry energy $\beta$, the slope parameter $L$ of the symmetry energy, the $\kappa$ parameter, and so on.
In particular, if a strong correlation between $\kappa$ and $\Delta r_{\rm{np}}$ could be established, then one can use the latter from experiments like the parity-violating electron scattering experiments (PREX, CREX)\,\cite{Abr12,Hor12,Adh21,Adh22,Rei22,Yuk22,Zha22,HuB22} to effectively constrain $\kappa$.
However, before the lower order parameters such as the surface symmetry energy $\beta$ are well constrained, it is hard to investigate the real effects of $\kappa$, since $\beta$ is directly related to the $\theta$ parameter as $\theta\sim\beta^{-1}$.
Nonetheless the neutron skin thickness for neutron-rich nuclei may provide a promising probe to detect the $\kappa$ parameter, and this is left for future studies.

\section{Expression for the Sixth-order Symmetry Energy of Finite Nuclei}\label{SS6}

If one considers even higher order isospin dependent terms in the nuclear surface tension coefficient as $\sigma/\sigma_0\approx1+y+\kappa y^2+s y^3$ with $s$ being the effective parameter beyond $\kappa$, one can directly obtain the sixth-order symmetry energy $a_{\rm{sym},6}(A)$ of finite nuclei as a function only of $A$ including both the volume and surface contributions as
\begin{align}\label{ee6}
a_{\rm{sym,6}}(A)
=&\left(1+\frac{E_{\rm{sym}}(\rho_0)}{\beta A^{1/3}}\right)^{-6}\Bigg[E_{\rm{sym,6}}(\rho_0)
-L_{\rm{sym,4}}\left(\frac{L}{K_0}\right)\notag\\
&\hspace*{1.cm}+\frac{K_{\rm{sym}}}{2}\left(\frac{L}{K_0}\right)^2
-\frac{J_0}{6}\left(\frac{L}{K_0}\right)^3
\notag\\
&\hspace*{-1.cm}+\frac{16\theta^2E_{\rm{sym}}^6(\rho_0)}{\beta A^{1/3}}\left(\frac{4\kappa^2}{1+\beta A^{1/3}/E_{\rm{sym}}(\rho_0)}-s\right)
\Bigg],
\end{align}
here $K_{\rm{sym}}\equiv 9\rho_0^2\d^2E_{\rm{sym}}(\rho)/\d\rho^2|_{\rho=\rho_0}$ is the curvature coefficient (see, e.g., Ref.\,\cite{ZhouY19}) of the symmetry energy, $J_0\equiv 27\rho_0^3\d^3E_0(\rho)/\d\rho^3|_{\rho=\rho_0}$ is the skewness coefficient (see, e.g., Ref.\,\cite{Cai17}) of the EOS $E_0(\rho)$ of SNM, and $L_{\rm{sym,4}}\equiv 3\rho_0\d E_{\rm{sym},4}(\rho)/\d\rho|_{\rho=\rho_0}$ is the slope parameter of the fourth-order symmetry energy $E_{\rm{sym,4}}(\rho)$ (see, e.g., Ref.\,\cite{Che09}).
The surface contribution (the last line) in Eq.~(\ref{ee6}) comes from two parts: the higher-order term originating from the lower-order coefficient $\kappa$ (proportional to $\kappa^2$) and the higher-order term from the coefficient $s$.
Naturally the surface term of $a_{\rm{sym},6}(A)$ approaches zero as $A\to\infty$.

Similarly to the investigations given in the last section, it is very difficult to extract the value of the sixth-order symmetry energy $E_{\rm{sym,6}}(\rho_0)\equiv 720^{-1}\partial^6E(\rho_0,\delta)/\partial\delta^6|_{\delta=0}$ for infinite matter from this expression even if the $a_{\rm{a,6}}$ appeared in the mass formula via the term $a_{\rm{a,6}}I^6$ is constrained, since one could adjust the coefficient $s$ to make $E_{\rm{sym,6}}(\rho_0)$ small or large without essentially affecting the properties of finite nuclei, such as the nuclear surface tension.
For example, without including the surface term here a ``small'' $a_{\rm{a,6}}\approx1\,\rm{MeV}$ (which can hardly be ``probed'' because of the tiny factor $I^6$ associated with it) may induce an $E_{\rm{sym,6}}(\rho_0)$ of about 7-8\,MeV, which is likely to be in What implication does a
sizableconflict with the empirical constraint on the EOS of ANM.
In this sense, the surface contribution to $a_{\rm{sym},6}(A)$ is fundamental.
In addition, by requiring, e.g., $E_{\rm{sym},6}(\rho_0)\lesssim1\,\rm{MeV}$ and $a_{\rm{sym},6}(A)\lesssim1\,\rm{MeV}$, Eq.\,(\ref{ee6}) then provides a link relating several characteristics with sizable magnitude, which could be used to establish certain correlations among them.
These are left for future studies.

\section{Summary and conclusions}\label{S4}

We have shown that for the fourth-order symmetry energy of finite nuclei, it could be naturally decomposed into two terms characterizing the bulk and the surfaces contributions, respectively.
The surface contribution $4\theta\kappa E_{\rm{sym}}^4(\rho_0)/\beta A^{1/3}/(1+E_{\rm{sym}}(\rho_0)/\beta A^{1/3})^4\sim4\theta\kappa E_{\rm{sym}}^4(\rho_0)/\beta A^{1/3}$ characterized by the product of $\theta\kappa$, to the fourth-order symmetry energy of finite nuclei is obtained by introducing the next leading order contribution to the isospin dependence of the nuclear surface tension coefficient through $\sigma/\sigma_0\approx1-\theta\mu_{\rm{a}}^2+\kappa\theta^2\mu_{\rm{a}}^4$, where $\kappa$ is an effective parameter characterizing the high order effects.
Although the $\kappa$ parameter may induce a sizable $4\theta\kappa E_{\rm{sym}}^4(\rho_0)/\beta A^{1/3}$, it has very little impact on the nuclear structure quantities of interest such as the surface tension itself (since $y=-\theta\mu_{\rm{a}}^2$ is generally small than unity).
Since this term is independent of the $E_{\rm{sym,4}}(\rho_0)$, it characterizes the coupling between the symmetry energy $E_{\rm{sym}}(\rho_0)$ and the high order coefficient $\kappa$, i.e., it is induced by some high-order isospin dependent surface tension effects.
Although both the surface contribution and the bulk term to the fourth-order symmetry energy of finite nuclei could be large, they contribute little to the nuclear structure quantities, since generally one has $I^4\lesssim0.003$ for finite nuclei. This means that the fourth-order symmetry energy of finite nuclei is usually hard to ``probe'' and special observables are needed (see, e.g., Refs.\,\cite{Jia14,Jia15,Tia16}).

On the other hand, all the microscopic many-body theories and phenomenological model predictions give the consistent constraint that $E_{\rm{sym,4}}(\rho_0)\lesssim2$\,MeV, indicating the $E_{\rm{sym,4}}(\rho_0)$ could not be large although its counterpart of finite nuclei could be sizable.
One needs to consider the total fourth-order symmetry energy of finite nuclei composed of both the bulk and the surface contributions. In this sense the surface contribution characterized by $\kappa$ to the fourth-order symmetry energy of finite nuclei is fundamental, i.e., it is essential for explaining a reasonable $E_{\rm{sym,4}}(\rho_0)\lesssim2\,\rm{MeV}$ and a sizable $a_{\rm{sym,4}}(A)$ simultaneously.
Essentially, it is hard to constrain the parameter $\kappa$ via finite-nucleus information.
In the future, unless certain quantities/processes determining the coefficient $\kappa$ to within some narrow range are available, it seems that one can hardly constrain the fourth-order symmetry energy $E_{\rm{sym,4}}(\rho_0)$ of nuclear matter from the nuclear mass formula fitting on $a_{\rm{sym,4}}(A)$, since a finite nucleus has a surface while the infinite matter does not.

Finally, we also present the expression for the sixth-order symmetry energy of finite nuclei, which is related to more nuclear matter bulk parameters and the higher-order isospin-dependent surface tension.
We would like to mention that
although the higher-order symmetry energies of finite nuclei are difficult to measure in terrestrial nuclei, they could be potentially useful for understanding the properties of neutron star crust or supernova explosions where extremely neutron-rich clusters may exist.

\section*{Acknowledgments}
This work was supported in part by the National Natural Science Foundation of China
under Grants No. 12235010, No. 11905302, and No. 11625521, the National SKA Program of China No. 2020SKA0120300, and the Fundamental Research Funds for the Central Universities, Sun Yat-Sen University (No. 22qntd1801).


\begin{references}
\bibitem{Zha01} F.S. Zhang and L.W. Chen, Chin. Phys. Lett. \textbf{18}, 142 (2001).
\bibitem{Ste06} A.W. Steiner, Phys. Rev. C \textbf{74}, 045808 (2006).
\bibitem{Xu09} J. Xun, L.W. Chen, B.A. Li, and H.R. Ma, Phys. Rev. C \textbf{79}, 035802 (2009); Astrophys. J. \textbf{697}, 1549 (2009).
\bibitem{Cai12} B.J. Cai and L.W. Chen, Phys. Rev. C \textbf{85}, 024302 (2012).
\bibitem{Sei14} W.M. Seif and D.N. Basu, Phys. Rev. C \textbf{89}, 028801 (2014).
\bibitem{Gon17} C. Gonzalez-Boquera, M. Centelles, X. Vinas, and A. Rios, Phys. Rev. C \textbf{96}, 065806 (2017).
\bibitem{PuJ17} J. Pu, Z. Zhang, and L.W. Chen, Phys. Rev. C \textbf{96}, 054311 (2017).

\bibitem{Lat91}J. M. Lattimer, C. J. Pethick, M. Prakash, and P. Haensel, Phys. Rev. Lett. \textbf{66}, 2701 (1991).
\bibitem{Yak01}D.G. Yakovlev, A.D. Kaminker, O.Y. Gnedin, and P. Haensel, Phys. Rep. \textbf{354}, 1 (2001).
\bibitem{Yak04}D.G. Yakovlev and C.J. Pethick, Annu. Rev. Astron. Astrophys. \textbf{42}, 169 (2004).

\bibitem{Dan02} P. Danielewicz, R. Lacey, and W.G. Lynch, Science,
\textbf{298}, 1592 (2002).

\bibitem{ditoro} V. Baran, M. Colonna, V. Greco, and M. Di Toro, Phys. Rep. \textbf{410}, 335 (2005).

\bibitem{LCK08} B.A. Li, L.W. Chen, and C.M. Ko, Phys. Rep. \textbf{464}, 113 (2008).

\bibitem{Tesym} B.A. Li, \`{A}. Ramos, G. Verde, I. Vida\~{n}a (Eds.), \textit{Topical issue on nuclear symmetry energy}, Euro. Phys. J. A \textbf{50}, No.2 (2014).


\bibitem{Col14} G. Colo, U. Garg, and H. Sagawa, Eur. Phys. J. A \textbf{50}, 26 (2014).

\bibitem{Bal16} M. Baldo and G.F. Burgio, Prog. Part. Nucl. Phys. \textbf{91}, 203 (2016).

\bibitem{Oer17} M. Oertel, M. Hempel, T. Klahn, and S. Typel, Rev. Mod. Phys.  \textbf{89}, 015007 (2017).

\bibitem{Garg18} U. Garg and G. Col\`{o}, Prog. Part. Nucl. Phys. \textbf{101}, 55 (2018).

\bibitem{LiBA18}B.A. Li \textit{et al.}, Prog. Part. Nucl. Phys. \textbf{99}, 29 (2018).

\bibitem{CaiLi2022} B.J. Cai and B.A. Li, Phys. Rev. C \textbf{105}, 064607 (2022).

\bibitem{CaiLi2022a}B.J. Cai and B.A. Li, Ann. Phys. \textbf{444}, 169062 (2022).

\bibitem{Che09}L.W. Chen \textit{et al.}, Phys. Rev. C \textbf{80}, 014322 (2009).
\bibitem{Agr17}B.K. Agrawal, S.K. Samaddar, J.N. De, C. Mondal, S. De, Int. J. Mod. Phys. E \textbf{26}, 1750022(2017).

\bibitem{Lee98} C.H. Lee, T.T. S. Kuo, G.Q. Li, and G. E. Brown, Phys. Rev. C \textbf{57}, 3488 (1998).
\bibitem{Bom91}I. Bombaci and U. Lombardo, Phys. Rev. C \textbf{44}, 1892 (1991).
\bibitem{Kai15}N. Kaiser, Phys. Rev. C \textbf{91}, 065201 (2015).


\bibitem{Mye69} W.D. Myers and W.J. Swiatecki, Ann. Phys. \textbf{55}, 395 (1969).

\bibitem{Brack85}M.Brack, C. Guet, and H. Hakansson, Phys. Rep. \textbf{123}, 275 (1985).


\bibitem{Mye96} W.D. Myers and W.J. Swiatecki, Nucl. Phys. \textbf{A601}, 141 (1996).


\bibitem{Dan03} P. Danielewicz, Nucl. Phys. \textbf{A727}, 233 (2003).


\bibitem{AME2012} M. Wang \textit{et al.}, Chin. Phys. C \textbf{36}, 1603 (2012).

\bibitem{Jia14} H. Jiang \textit{et al.}, Phys. Rev. C \textbf{90}, 064303 (2014).


\bibitem{Tia16} J.L. Tian, H.T. Cui, T. Gao, and N. Wang, Chin. Phys. C \textbf{40}, 094101 (2016).
\bibitem{Jia15} H. Jiang \textit{et al.}, Phys. Rev. C \textbf{91}, 054302 (2015).


\bibitem{Dan09}P. Danielewicz and J. Lee, Nucl. Phys. \textbf{A818}, 36 (2009).

\bibitem{Dan14} P. Danielewicz and J. Lee, Nucl. Phys. \textbf{A922}, 1 (2014).

\bibitem{WangRui17} R. Wang and L.W. Chen, Phys. Lett. \textbf{B773}, 62 (2017).

\bibitem{LiBA13} B.A. Li and X. Han, Phys. Lett. \textbf{B727}, 276 (2013).
\bibitem{Zha13} Z. Zhang and L.W. Chen, Phys. Lett. \textbf{B726}, 234 (2013).

\bibitem{You99}D.H. Youngblood, H.L. Clark, and Y.-W. Lui, Phys. Rev. Lett. \textbf{82}, 691 (1999).
\bibitem{Shl06}S. Shlomo, V.M. Kolomietz, and G. Colo, Eur. Phys. J. A \textbf{30}, 23 (2006).
\bibitem{Che12} L.W. Chen and J.Z. Gu, J. Phys. G \textbf{39}, 035104 (2012).

\bibitem{Cai21}B.J. Cai and B.A. Li, Phys. Rev. C \textbf{103}, 034607 (2021).

\bibitem{Chen17}L.W. Chen, Nucl. Phys. Rev. \textbf{34}, 20 (2017).
\bibitem{LiBA21}B.A. Li \textit{et al.}, Universe \textbf{7}, 182 (2021).

\bibitem{Abr12} S. Abrahamyan \textit{et al.}, Phys. Rev. Lett. 108, 112502 (2012).
\bibitem{Hor12}C.  Horowitz \textit{et al.}, Phys. Rev. C \textbf{85}, 032501(R) (2012).
\bibitem{Adh21}D. Adhikari \textit{et al.}, Phys. Rev. Lett. \textbf{126}, 172502 (2021).
\bibitem{Adh22}D. Adhikari \textit{et al.}, Phys. Rev. Lett. \textbf{129}, 042501 (2022).
\bibitem{Rei22}P.-G. Reinhard, X. Roca-Maza, and W. Nazarewicz, arXiv:2206.03134 (2022).
\bibitem{Yuk22}E. Yuksel and N. Paar, arXiv:2206.06527 (2022).
\bibitem{Zha22}Z. Zhang and L.W. Chen, arXiv:2207.03328 (2022).
\bibitem{HuB22}B.S. Hu \textit{et al.}, Nat. Phys. \textbf{18}, 1196 (2022).

\bibitem{ZhouY19}Y. Zhou and L.W. Chen, Astrophys. J.  \textbf{886},  52 (2019).
\bibitem{Cai17}B.J. Cai and L.W. Chen,  Nucl. Sci. Tech. \textbf{28}, 185 (2017).



\end{references}
\end{document}